\documentstyle[11pt,newpasp,twoside,epsf]{article}

\begin{document}

\title{The Quest for the Dominant Stellar Population in the Giant Elliptical NGC~5018}

\author{Lucio M. Buson$^1$, Francesco Bertola$^2$, David Burstein$^3$\\ and
Michele Cappellari$^2$}
\affil{$^1$Osservatorio Astronomico di Capodimonte, Napoli, Italy}
\affil{$^2$Dipartimento di Astronomia, Universit\`a di Padova, Padova, Italy}
\affil{$^3$Dept. of Physics \& Astronomy, Arizona State Univ., Tempe, AZ, USA}

\markboth{L.M. Buson et al.}{The Quest for the Dominant Stellar Population of the Giant Elliptical NGC~5018}

\section{Introduction}

NGC~5018, a relatively near, giant E3 galaxy, deserved quite a large attention in the past. Evidence of past interaction with other systems comes mainly from the extended pattern of dust lanes/patches clearly recognizable from ground-based observations. What is more, this object captures immediately the attention of the observer for (i) its anomalous location on the Mg$_2$-luminosity relation of Schweizer {\em et al.} (1990), having one of the weakest absorption-line strengths (Mg$_2$=0.218) relative to its absolute magnitude and central velocity dispersion among giant ellipticals and (ii) the extremely weak far-UV emission Bertola {\em et al.} (1993) find in its IUE spectrum (strictly resembling the UV energy distribution of the dwarf elliptical M32). Interesting enough, though one expects ongoing star formation plays a major role in a recent merger such as NGC~5018, both the observed low value of its optical Mg$_2$ line strength index {\em and} its weak far-UV flux can fit into scenarios {\em not} invoking recent star formation activity. According to this latter view, NGC~5018 has been alternatively described either as a sort of ``Celacanthus galaxy'', {\em i.e.} a rare, chemically unevolved giant elliptical hosting a dominating M32-like population (Bertola {\em et al.} 1993), or a system fully dominated by an intermediate-age (2.8~Gyr) population with a roughly solar composition as put forward by Leonardi \& Worthey (2000). On the other side, since other groups ({\em e.g.} Carollo \& Danziger 1994) are prone to conclude that a large amount of truly young population {\em and strong dust absorption} might conspire to cause the above UV/optical properties, strict constraints on the amount of internal extinction are urgently needed in order to properly address the issue of the dominant population age and metallicity.

\section{Data Analysis and Discussion}

The above goal can now be achieved thanks to the recent set of Schweizer's WFPC2 F555W and F814W deep images (program ID~7468) which allows us to finally get rid of the persistent suspicion of a {\em heavy} dust absorption affecting the UV emission of this galaxy. In particular, starting from the high S/N E(V$-$I) map shown in Fig.~1, one derives---after removing the underlying V$-$I color gradient---a luminosity weighted average value E(B$-$V)$\sim$0.02 due to structured dust within the IUE large aperture (10''$\times$20''). This, in turn, assures that the collected UV flux is only marginally affected by the internal extinction due to dust (namely $\sim$0.06~dex at $\lambda$~1500~\AA).

\begin{figure}
\plotfiddle{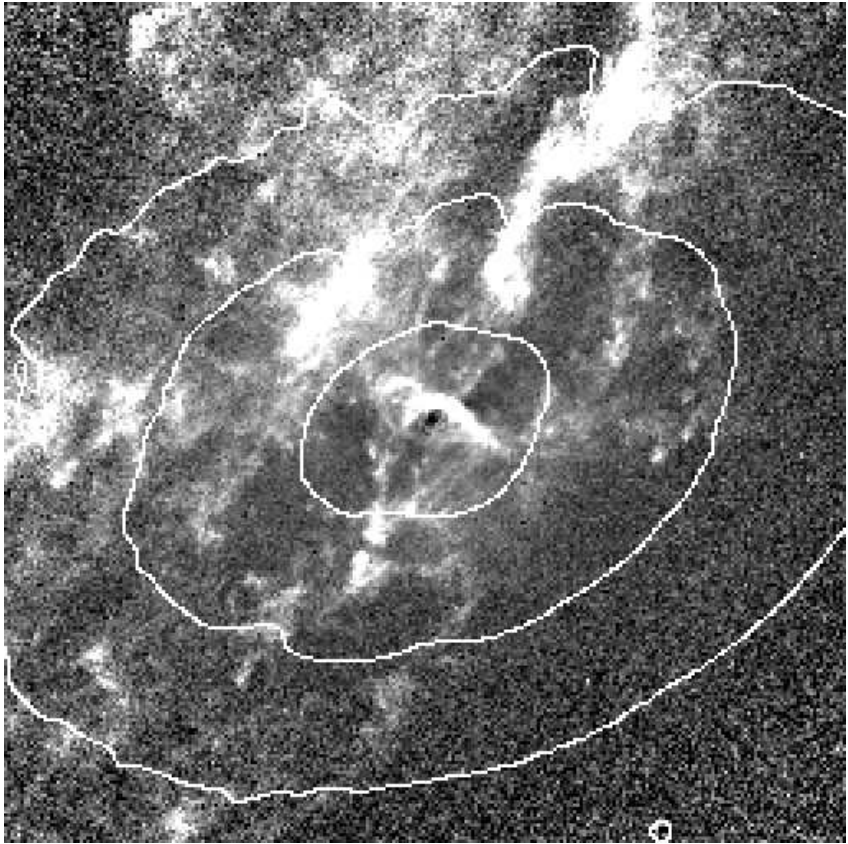}{10.6cm}{0}{86}{86}{-262}{-342}
\caption{Calibrated E(V$-$I) image of the inner portion (18''$\times$18''; $\sim$ 2$\times$ the IUE aperture) of the PC field of NGC~5018 showing its intricate net of dust lanes/patches. The brightest areas correspond to a maximum reddening value E(V$-$I)$\sim$0.35. The (V$-$I) color of the barely resolved nucleus is 0.3 mag bluer than the underlying stellar population. North is up and East to the left. A set of the F555W image isophotes is superimposed to help locating dust features.}
\end{figure}

\end{document}